\UseRawInputEncoding
\documentclass[twocolumn]{revtex4}
\usepackage{epsfig,graphics,subfigure,amssymb,amsmath,subeqnarray,color,xcolor,amssymb}

\begin{document}
\def\u{{\bf u}}\def\n{{\bf n}}\def\f{{\bf f}}
\def\t{{\bf t}}\def\d{{\rm d}}
\def\r{{\bf r}}\def\x{{\bf x}}
\def\e{{\bf e}}\def\1{{\bf 1}}\def\0{{\bf 0}}
\def\p{{\partial}}
\def\v{\vspace {2cm}}
\def\textheight{25cm}

\title{Calculus of variations  rationalises long-wavelength flagellar waves}

\title{Hydrodynamically optimal long-wavelength flagella have constant-speed travelling waves}

\title{Using calculus of variations to  rationalise long-wavelength flagellar waves}

\title{Hydrodynamically optimal long-wavelength flagella deform as  travelling waves}

\title{Travelling waves are hydrodynamically optimal for long-wavelength flagella}

\author{Eric Lauga}
\email{e.lauga@damtp.cam.ac.uk}
\affiliation{Department of Applied Mathematics and Theoretical Physics, 
University of Cambridge, Wilberforce Road, Cambridge CB3 0WA, UK}
\date{\today}
\begin{abstract}
Swimming eukaryotic microorganisms such as spermatozoa, algae and ciliates   self-propel in viscous fluids using travelling wave-like deformations of slender appendages called flagella. Waves are predominant because Purcell's scallop theorem precludes time-reversible kinematics for locomotion.  
Using the calculus of variations on a periodic long-wavelength model of flagellar swimming, we show that the planar flagellar kinematics maximising  the time-averaged propulsive force for a fixed amount of energy dissipated in the surrounding   fluid correspond for all times to  waves travelling  with constant speed, potentially on a curved  centreline, with propulsion always in the direction opposite to the wave.

\end{abstract}
\maketitle

\section{Introduction}
 
A large proportion of all micro-organisms are able to self-propel in liquids, including spermatozoa, algae, ciliates and many species of bacteria~\cite{yates86}. In order to achieve self-propulsion, most cells exploit flagella -- slender whiplike organelles extending out of the cell body and whose time-varying deformations powered by molecular motors induces flows and creates the propulsive forces required to overcome viscous friction~\cite{braybook}. Due to their small sizes and speeds, swimming cells move   at low Reynolds number, a regime which is qualitatively different from the high-Reynolds world of fish and birds, and with well-studied physical  consequences~\cite{childress81}.  A lot of   work has been devoted by   the scientific community to uncover  the physical principles behind the dynamics of individual cells and to capture  their interactions in complex environments~\cite{gray68,lighthill75,lighthill76,brennen77,lp09}.

 Despite the tremendous variation in size, geometry, kinematics, ecosystems and evolutionary position on the tree of life,  one universal feature of swimming cells is the use of waves to induce locomotion. The physical rationale for the appearance of travelling waves lies in Purcell's famed scallop theorem, which states that the time-reversible motion of a swimmer can never produce locomotion~\cite{purcell77}. Therefore, biological swimmers need  to deform in a manner that indicates a clear  direction of time, the prototypical example of which is a wave. Bacteria use helical semi-rigid filaments, each  of which is rotated by a specialised rotary motor leading to apparent longitudinal waves of displacement \cite{berg00_PT}. The physics of helical propulsion is now well understood~\cite{lauga16}, including the fact that the flagellar shapes 
observed in  bacteria   are close to that  of the  optimal propeller predicted by theory~\cite{spagnolie2011}. 

 Swimming eukaryotic  cells (i.e.~biological cells with a nucleus, in contrast  with bacteria) also use flagellar waves in order to self-propel. Unlike   bacteria, the flagella of eukaryotes are flexible, active filaments that deform with travelling wave-like motion, in  most cases planar.  
 While a  bacterial flagellum is passively rotated by a single motor, the actuation of  eukaryotes is spatially distributed inside  the flagellum and it involves the coordinated action of a large number  of molecular motors~\cite{julicher1997modeling}. The model organisms   from which we have learnt the most include   (i) spermatozoa, mostly of invertebrates and  mammals~\cite{fauci06,gaffney11}; (ii) 
 single cells such as the algal genus {\it Chlamydomonas} equipped with a pair of flagella~\cite{pedley92,stocker} or ciliates, e.g.~the genus {\it Paramecium}~\cite{blake74,brennen77}; and (iii) 
     multicellular organisms such as the algal genus {\it Volvox} powered by a large number of short flagella~\cite{goldstein:15}.
   
All these organisms have   in common that their swimming is propelled by travelling wave-like motion of their  flagella. Such a simple observation within such a wide biological spectrum has invariably lead the scientific  community to look for flagellar waving as the solution to an optimisation problem. The most formal result was obtained by Pironneau and  Katz using the calculus of variations~\cite{pironneau74}, a well known mathematical framework to address problems in shape optimisation, with broad applications in fluid mechanics at both high~\cite{mohammadi2004shape} and low Reynolds numbers~\cite{pironneau1974optimum,montenegro2015other}.  {In their   paper, Pironneau and Katz considered   organisms  with  no cell body swimming using the two-dimensional waving of a   slender flagellum.  {Using the resistive-force theory of slender filaments,} they showed that the
 kinematics of  the deforming flagellum that lead to locomotion with the least amount of energy dissipated in the surrounding fluid,  subject to   a fixed constraint of  local flagellum velocity projected on the  tangent  of the  flagellum and spatially-averaged,  correspond in the swimming frame to an   apparent sliding of the flagellum along its instantaneous centreline. This  suggested  flagellar waves travelling instantaneously in the direction opposite to that of the locomotion~\cite{pironneau74}.}

This energetic advantage  of flagellar waves encouraged the scientific community to then hunt for the optimal wave.  In two dimensions, the hydrodynamically optimal waving sheet has to be characterised numerically and takes the form of a  
front-back symmetric regularised cusp~\cite{montenegro2014optimal}.
For a  slender filament, the optimal wave has a constant angle between the local tangent to the filament and the  direction of   wave propagation taking therefore the shape of a helix for three-dimensional motion and straight segments joined in kinks for planar flagellar  waves~\cite{pironneau74,pironneau_inbook,lighthill75}. This kink can be regularised by taking into account the elastic cost of bending the flagellum~\cite{Spagnolie2010} or the irreversible nature of the internal rate of working of the molecular motors powering the flagellum~\cite{laugaeloy13}. Optimal waving motion has also been investigated for  biflagellate algae~\cite{tam2011optimal},  simple swimmers with small numbers of degrees of freedom~\cite{tam07} 
and   coordinated waving in the  locomotion and feeding of ciliates~\cite{michelin2010,michelin2011,osterman2011finding,eloylauga12,michelin2013}. More formal mathematical approaches to the optimisation problem have also been developed~\cite{avron04:opt,alouges08,AlougesDeSimoneLefebvre2009}. Recently, reinforcement learning has been proposed as a mechanism to help design favourable swimming strokes  for simple artificial swimmers~\cite{tsang2020self}.

 In this paper, we show that a simple calculation using the tools of the calculus of variations allows to rationalise analytically the predominance of constant-speed flagellar waves for long-wavelength motion.  Assuming planar motion, we mathematically determine the flagellar kinematics maximising the time-averaged propulsive force for a fixed amount of total energy dissipated in the surrounding fluid. We show that they correspond, for all times, to  waves travelling  with constant speed, potentially on a curved flagellum centreline, and always lead to propulsion in the direction opposite to that of the wave.

      \begin{figure}[t]
   	\includegraphics[width=0.5\textwidth]{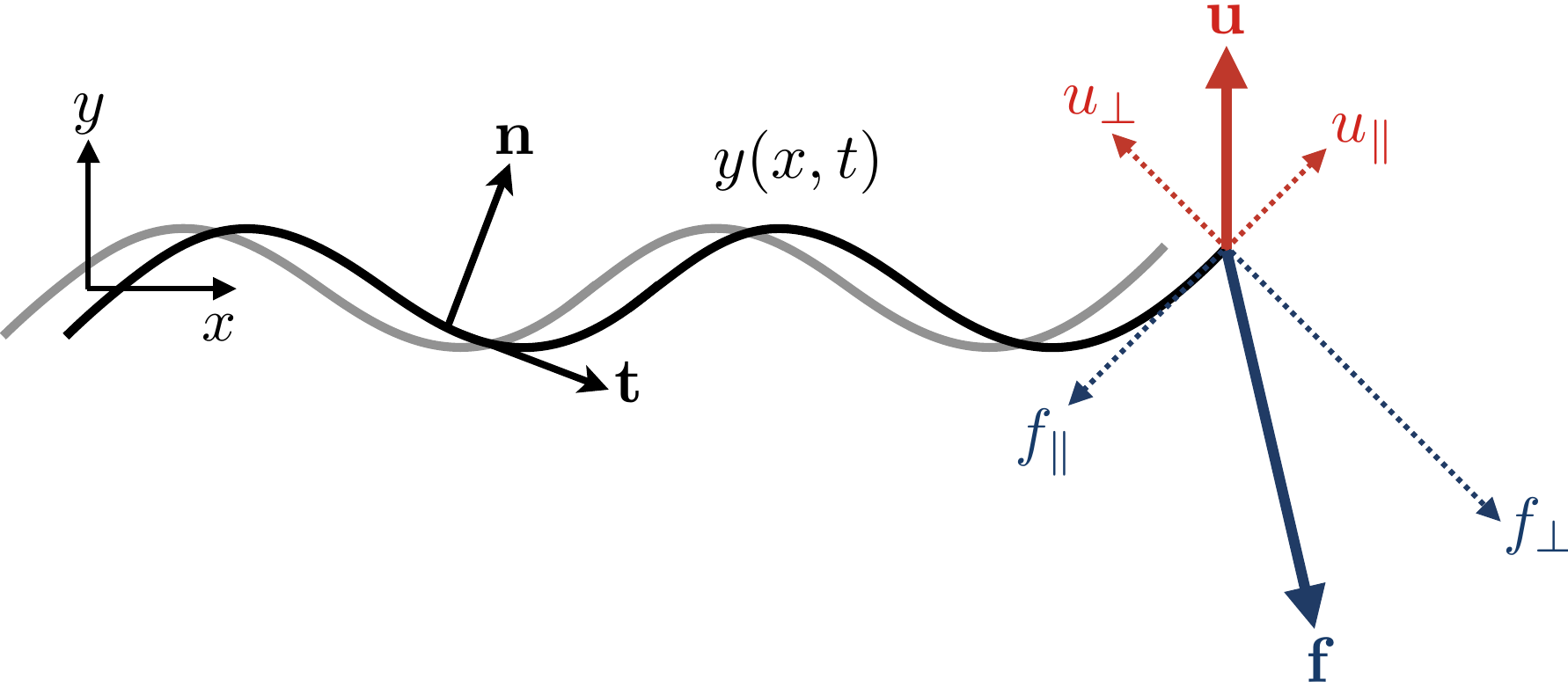}
   	\caption{An infinite flexible flagellum undergoing planar beating is described mathematically by a time-varying   function $y(x,t)$ in Cartesian coordinates. The unit vectors $\t$ and $\n$ are the local  tangent and normal to the instantaneous flagellar shape. The instantaneous velocity of the flagellum centreline, $\u$, can be decomposed into components along the directions tangent ($u_\parallel$) and normal to the flagellum ($u_\perp$), and similarly for the hydrodynamic force density acting on the flagellum, $\f$	(components $f_\parallel$ and $f_\perp$). The force and velocity components are related to one another through resistive-force theory. 
	We use the calculus of variations to determine the class of functions $y(x,t)$ leading to the maximal average propulsive force for  a given amount of average energy dissipated in the fluid.
}
	\label{fig1}
   \end{figure}

\section{Propulsion and energetics of a deforming  flagellum}
\subsection{Setup}

Locomotion is induced in a Newtonian fluid  by the planar deformation of 
a  flexible inextensible flagellum (setup in Fig.~\ref{fig1}).  The shape of the  flagellum is described in Cartesian coordinates by material points  $\big(x,y(x,t)\big)$, and the flagellum is assumed to be infinite and slender. We focus on the limit of long-wavelength motion and therefore we assume  $|\p y/\p x |\ll 1$ everywhere. The flagellum is  spatially-periodic with fixed period $\lambda$ in the $x$ direction and the deformations are periodic in time with fixed period $T$.

\subsection{Propulsive force}

We first compute the propulsive force generated by the deforming flagellum, i.e.~the net hydrodynamic force experienced by the total flagellum along the swimming direction (assumed to be $x$) and resulting from its time-varying motion. Denoting by $\u$ the velocity of the flagellum centreline relative to the background fluid, we use the resistive-force theory of viscous hydrodynamics, an asymptotic result
valid in the limit of slender filaments, to compute the hydrodynamic forces acting on the filament in the limit of low Reynolds number~\cite{cox70,batchelor70}. 

In this limit, the local hydrodynamic force per unit length along the filament, $ \f$,  is decomposed into its component parallel to the local unit tangent vector $\t$, $\f_\parallel = ( \f\cdot\t)\t$, and its    perpendicular component, $\f_\perp = \f -\f_\parallel =( \f\cdot\n)\n $,
where  $\n$ is the unit normal to the flagellum chosen so that $\t\times \n = \e_z$. The magnitudes of these two force components are 
\begin{equation}\label{eq:twocomponents}
f_\parallel =-c_\parallel u_\parallel,\quad 
f_\perp = -c_\perp u_\perp,
\end{equation}
where $u_\parallel$ and $u_\perp$ are the local components of the flagellum velocity parallel and perpendicular to the local tangent  (see sketch in Fig.~\ref{fig1}).  {In Eq.~\eqref{eq:twocomponents}, the parameters} 
$c_\parallel$ and $c_\perp$  are the perpendicular and parallel drag coefficients, given in the slender limit $a\ll L$ by $c_\perp\approx 2 c_\parallel\approx 4\pi\mu/\ln(L/a)$ where 
$a$ is  the cross sectional radius of the flagellum and  $L$ is either its length if the flagellum is finite~\cite{cox70,batchelor70} or a characteristic length proportional to the wavelength if it is infinite~\cite{lighthill75}.

The two components in Eq.~\eqref{eq:twocomponents} can be combined to obtain an equation relating   the   hydrodynamic force density vector, $ \f$, 
to the local filament velocity vector, $\u$, 
 as
 \begin{equation}\label{eq:rft}
 \f = -[c_\parallel \t\t + c_\perp \n\n]\cdot\u.
\end{equation}

The $x$-component of the time-averaged propulsive force, denoted by $F_x $,  is then given as a double  integral over the period and the wavelength
\begin{equation}\label{eq:Fx_first}
F_x = \frac{1}{T}\int_0^T\int_0^\lambda  \f\cdot \e_x \,
\,\d x\,\d t,
\end{equation}
which, given Eq.~\eqref{eq:rft}, is equal to
\begin{equation}\label{eq:Fxint}
F_x =- \frac{1}{T}\int_0^T\int_0^\lambda  
[c_\parallel (\t\cdot\e_x)(\u\cdot \t) + c_\perp (\n\cdot\e_x)(\u\cdot \n)]
 \,\d x\,\d t.
\end{equation}
In the case of a flagellum where $|\p y/\p x |\ll 1$, we have the leading-order long-wavelength geometry and velocity given by
\begin{equation}\label{eq:geometry}
\t = \left(
\begin{array}{c}
  1   \\
   \displaystyle \frac{\p y}{\p x}    
\end{array}
\right),\quad \n= \left(
\begin{array}{c}
\displaystyle-\frac{\p y}{\p x}         \\
    1
\end{array}
\right),
\quad \u = \left(
\begin{array}{c}
  0 \\
\displaystyle    \frac{\p y}{\p t}    
\end{array}
\right),
\end{equation}
so that
\begin{equation}
\u\cdot \t = \frac{\p y}{\p x}  \frac{\p y}{\p t} ,\quad \u\cdot\n = \frac{\p y}{\p t},\end{equation}
and thus   the  integrand in Eq.~\eqref{eq:Fx_first} becomes
\begin{equation}
 \f\cdot \e_x = (c_\perp - c_\parallel) \frac{\p y}{\p x}  \frac{\p y}{\p t},
\end{equation}
at leading order in the long wavelength limit. 

Integrating in space over one wavelength and calculating the average in time leads to the mean propulsive force  as
\begin{equation}\label{eq:final_prop}
F_x 
= \frac{(c_\perp - c_\parallel)}{T}\int_0^T\int_0^\lambda
\frac{\p y}{\p t}\frac{\p y}{\p x}\,\d x\, \d t.
\end{equation}

We now wish to determine  the function $y(x,t)$ of both time and space that maximises the propulsive force in Eq.~\eqref{eq:final_prop} for a given amount of energy expended by the flagellum.   {In the long-wavelength approximation, the drag per unit length exerted on a flagellum translating with speed $U$ is equal to $-c_\parallel U  $ at leading order. Hence   the swimming speed of the flagellum in   Stokes flow is   directly proportional to the propulsive force with a coefficient of proportionality independent of the shape; maximising $F_x $ is thus equivalent to maximising $U$. }

\subsection{Flagellum energetics}
We next make the standard   assumption  that the majority of the energy expended by the flagellum is dissipated in the surrounding viscous fluid~\cite{lighthill76}, and thus neglect internal sources of dissipation. The rate of  energy dissipation in the fluid is equal to the 
 rate of working (or power) of the moving flagellum against the fluid. Since $\f$ is the hydrodynamic force per unit length acting on the filament, the power density is $\dot w = -  \f \cdot \u$, which we then need to integrate over the whole length of the flagellum. Given Eq.~\eqref{eq:rft}, we have
\begin{equation}
\dot w = c_\perp(\u\cdot \n)^2 + c_\parallel (\u\cdot \t)^2.
\end{equation}
Using the results of Eq.~\eqref{eq:geometry}, we see that the leading-order value of the  power density is given by 
\begin{equation}
\dot w = c_\perp \left(\frac{\p y}{\p t}\right)^2,
\end{equation}
plus small corrections in the long-wavelength limit. 
The leading-order time-averaged rate of working of the flagellum  over one wavelength, $\dot W$,   is therefore  given by the double integration in time and over a wavelength as
\begin{equation}\label{eq:dotW}
\dot W 
=\frac{c_\perp}{T}\int_0^T\int_0^\lambda  \left(\frac{\p y}{\p t}\right)^2\,\d x\, \d t.
\end{equation}
 Notably, we see that the dissipation is  dominated at leading order by the lateral motion of the flagellum.

\section{Hydrodynamically optimal active flagellum}

\subsection{Optimisation setup}

We   wish to determine the time-varying shape of the periodic flagellum that {maximises} the value of $F_x$ (Eq.~\ref{eq:final_prop}) for a given value of $\dot W$ (Eq.~\ref{eq:dotW}). For this   optimisation problem we define  the Lagrangian 
\begin{equation}\label{eq:Lagrangian}
{\cal L}[y] \equiv F_x + \Gamma (\dot W - \dot W_0),
\end{equation}
 where $\Gamma$ is a constant Lagrange multiplier and the constant $\dot W_0$ is the value of the constraint for $\dot W$. By considering a small periodic change in the shape of the flagellum, $y\to y + \delta y$, we now calculate the resulting change in the functional,  ${\cal L} \to {\cal L} + \delta {\cal L}$. We will then ask what time-varying flagellum shape renders the first functional derivative zero, i.e.~$\delta {\cal L} = 0$. Throughout we assume that the flagellum is periodic in space with wavelength $\lambda$ and periodic in time with period $T$, both of which are constant, so that we require 
\begin{equation}\label{eq:periodicity}
y(x+\lambda,t)=y(x,t),\quad y(x,t+T)=y(x,t).
\end{equation}

\subsection{Calculus of variations}

To compute the change in ${\cal L}$ we write $\delta {\cal L} = \delta F_x + \Gamma \delta\dot W$ and calculate each term separately.

\subsubsection{Variation of the propulsive force}
Since the propulsive force is quadratic in the flagellum shape and its velocity, a perturbation of Eq.~\eqref{eq:final_prop} leads to
\begin{equation}
\delta F_x = \frac{(c_\perp - c_\parallel)}{T}\int_0^T\int_0^\lambda
\left(
\frac{\p y}{\p t}\frac{\p \delta y}{\p x}
+\frac{\p \delta  y}{\p t}\frac{\p y}{\p x}
\right)
\,\d x\, \d t.
\end{equation}
We then  use integration by parts to evaluate each  term. For the first one we write
\begin{eqnarray}\label{eq:deltaFx}
\int_0^T\int_0^\lambda\notag
\frac{\p y}{\p t}\frac{\p \delta y}{\p x}\,\d x\, \d t
=\int_0^T\int_0^\lambda
\frac{\p}{\p x}\left(\frac{\p y}{\p t}
 \delta y\right)\,\d x\, \d t\\
-\int_0^T\int_0^\lambda
 \delta y \frac{\p^2 y}{\p t\p x}\,\d x\, \d t.
\end{eqnarray}
The spatial integral of the term $\frac{\p}{\p x}\left(\frac{\p y}{\p t}
 \delta y\right)$ is zero by  spatial periodicity of the flagellum velocity and   its perturbation, and thus
\begin{equation}
\int_0^T\int_0^\lambda
\frac{\p y}{\p t}\frac{\p \delta y}{\p x}
\,\d x\, \d t= -
\int_0^T\int_0^\lambda
\delta y \frac{\p^2 y}{\p t\p x}
\,\d x\, \d t.
\end{equation}
The same manipulation can be carried out for the second term on the right-hand side of Eq.~\eqref{eq:deltaFx} and we find 
\begin{eqnarray}\label{eq:17}
\int_0^T\int_0^\lambda\notag
\frac{\p y}{\p x}\frac{\p \delta y}{\p t}\,\d x\, \d t
=\int_0^T\int_0^\lambda
\frac{\p}{\p t}\left(\frac{\p y}{\p x}
 \delta y\right)\,\d x\, \d t\\
-\int_0^T\int_0^\lambda
 \delta y \frac{\p^2 y}{\p t\p x}\,\d x\, \d t,
\end{eqnarray}
and this time the first integral on the right-hand side is zero by periodicity in time of the flagellum and its perturbation.  We therefore obtain the final result for the variation of the force
\begin{equation}\label{eq:Ffinal}
\delta F_x = \frac{2(  c_\parallel - c_\perp)}{T}\int_0^T\int_0^\lambda
 \delta y \frac{\p^2 y}{\p t\p x}\,\d x\, \d t.
 \end{equation}
 
\subsubsection{Variation of the rate of working}

The perturbation of the expression for the rate of working is  dealt with similarly. Here, again, we have a quadratic quantity, so perturbing Eq.~\eqref{eq:dotW} leads to
\begin{equation}\label{eq:Wbefore}
\delta W = \frac{2c_\perp}{T}\int_0^T\int_0^\lambda 
\frac{\p y}{\p t}\frac{\p \delta y}{\p t}
 \,\d x\, \d t.
\end{equation}
Using an integration by parts in time and invoking again periodicity in time we obtain 
\begin{equation}\label{eq:Wfinal}
\delta W = -\frac{2c_\perp}{T}\int_0^T\int_0^\lambda 
\frac{\p^2 y}{\p t^2} \delta y 
 \,\d x\, \d t.
\end{equation}

\subsubsection{Optimal waveform}

Given Eqs.~\eqref{eq:Ffinal} and \eqref{eq:Wfinal}, we obtain the variation in the Lagrangian as
\begin{equation}
\delta {\cal L} = 
\frac{2}{T}\int_0^T\int_0^\lambda 
\delta y \left[
(c_\parallel - c_\perp)
\frac{\p^2 y}{\p t\p x}
-\Gamma c_\perp\frac{\p^2 y}{\p t^2} 
\right] \,\d x\, \d t.
\end{equation}

The standard first-order optimality condition is $\delta {\cal L} =0$ for all perturbations $\delta y$ and therefore the waveform shape is governed by  the partial differential equation
\begin{equation}\label{eq:optPDE}
\Gamma c_\perp\frac{\p^2 y}{\p t^2} =(c_\parallel - c_\perp)
\frac{\p^2 y}{\p t\p x}
\cdot
\end{equation}
This equation can be integrated in time  once to obtain
\begin{equation}\label{eq:optcond}
\Gamma c_\perp\frac{\p y}{\p t} =(c_\parallel - c_\perp)
\frac{\p y}{\p x}+ A(x)
,
\end{equation}
where   $A(x)$ is an arbitrary periodic function. 
To simplify notation we write this function as 
$A(x)\equiv(c_\perp-c_\parallel ){\cal H}'(x)$ so that the optimality condition in Eq.~\eqref{eq:optcond} may be rewritten as 
\begin{equation}
\Gamma c_\perp\frac{\p y}{\p t}=(c_\parallel - c_\perp)
\left(\frac{\p y}{\p x}
-{\cal H}'(x)\right)
.
\end{equation}
We next define the new variable $z(x,t)\equiv y(x,t)-{\cal H}(x)$, for which we  obtain a {transport equation}   
\begin{equation}\label{eq:waveeq}
\frac{\p z}{\p t}
=\frac{(c_\parallel - c_\perp)}{\Gamma c_\perp}\frac{\p z}{\p x}.
\end{equation}
The solutions to Eq.~\eqref{eq:waveeq} are  waves travelling with constant speed. 
The energetically-optimal solution   breaks therefore the right-left symmetry of the problem, recovering what was postulated  phenomenologically for active filaments~\cite{goldstein2016elastohydrodynamic}. As a consequence, the general solution for the flagellum waveform $y(x,t)$ is given by
\begin{equation}\label{eq:wavefinal}
y(x,t) ={\cal H}(x)+{\cal Y}\left[x - \frac{(c_\perp- c_\parallel )}{\Gamma c_\perp} t \right],
\end{equation}
with both $\cal H$ and $\cal Y$  periodic functions of period $\lambda$. Importantly, these  long-wavelength functions are arbitrary, provided that they satisfy the periodicity (Sec.~\ref{sec:periodicity}) and energy constraints (Sec.~\ref{sec:energy}).  
 The flagellar kinematics 
   maximising the mean propulsive force $F_x$ for a fixed value of energy dissipated in the fluid  $\dot W_0$ are therefore travelling waves; the function $\cal H$ represents the average shape of the flagellum centreline, and $\cal Y$  the shape of the travelling wave around this average. In nature, many flagellated species have shapes  that   oscillate periodically around a straight configuration and correspond to the case ${\cal H }= 0$~\cite{brennen77}. Some organisms however have a non-straight centreline, for example  the spermatozoa of chinchilla~\cite{woolley03},   bull~\cite{friedrich2010high} and some insects~\cite{pak2012hydrodynamics}, all of which would correspond to the case where ${\cal H }\neq 0$.

   \subsubsection{Lagrange multiplier}
   \label{sec:periodicity}
  The value of the Lagrange multiplier is obtained as a consequence of the periodicity of the waveform. Since $y(x,t)$  satisfies the spatio-temporal periodicity  from Eq.~\eqref{eq:periodicity}, for the function    $\cal Y$   of a travelling-wave argument  in Eq.~\eqref{eq:wavefinal} to follow the same periodicity  
   it is necessary that 
  \begin{equation}
 \frac{(c_\perp- c_\parallel )}{\Gamma c_\perp} T=\pm  \lambda ,
\end{equation}
where $\lambda$ and $T$  are assumed to be the shortest periods in  $x$ and  $t$. As a consequence we obtain   \begin{equation}\label{eq:Gamma}
\Gamma= \pm \frac{(c_\perp- c_\parallel )}{ c_\perp} \frac{T}{\lambda}.
\end{equation}
The  Lagrange multiplier has dimensions of an inverse speed, as expected from the form of the Lagrangian in Eq.~\eqref{eq:Lagrangian}. Note that  since 
$c_\perp - c_\parallel > 0$, the case $\Gamma > 0$ in Eq.~\eqref{eq:wavefinal} corresponds to a wave travelling in the $+x$ direction, 
while $\Gamma < 0$ means that the wave is  travelling in the $-x$ direction.

   \subsubsection{Energetic constraint}
   \label{sec:energy}
   
   The other constraint for the functions in Eq.~\eqref{eq:wavefinal} are related to the energy requirement,  $\dot W=\dot W_0$. 
  In order to evaluate the rate of working from the flagellum 
    we use the fact that $\cal Y$ is a travelling wave, Eq.~\eqref{eq:wavefinal}, and denote by  $\eta\equiv x-(c_\perp-c_\parallel )t/(\Gamma c_\perp)$ its argument  to obtain
\begin{equation}\label{eq:misceqs1}
\frac{\p y}{\p t}= \frac{\partial \cal Y}{\partial t}=
\frac{( c_\parallel-c_\perp )}{\Gamma c_\perp}\frac{\d \cal Y}{\d \eta}.
\end{equation}
Substituting this result into Eq.~\eqref{eq:dotW}   and using Eq.~\eqref{eq:Gamma}  for the Lagrange multiplier we   obtain
\begin{equation}\label{eq:W0}
\dot W _0
=c_\perp\frac{ \lambda^2}{T^2  }\int_0^\lambda  
\left[\cal Y'(\eta)\right]^2
\,\d \eta.
\end{equation}
The energy constraint provides therefore an integral constraint for the long-wavelength function $\cal Y$  (Eq.~\ref{eq:W0}) but no constraint on the function $\cal H$ beyond its periodicity.

\subsubsection{Direction of propulsion}

In order to determine the direction of propagation of the travelling wave       we
 evaluate the sign of the   propulsive force in Eq.~\eqref{eq:final_prop}.  Given Eq.~\eqref{eq:wavefinal} we have
\begin{equation}
\frac{\p y}{\p x}=   \frac{\partial \cal Y}{\partial x} + \frac{\d \cal H }{\d x } ,
\end{equation}
and, together with Eq.~\eqref{eq:misceqs1}, this means that the force in  Eq.~\eqref{eq:final_prop} can be expressed as
\begin{equation}\label{eq:final_prop_2}
F_x 
= \frac{(c_\perp - c_\parallel)}{T}
\int_0^T\int_0^\lambda
\frac{\partial \cal Y}{\partial t}
\left(\frac{\partial \cal Y}{\partial x} + \frac{\d \cal H }{\d x } \right)
\,\d x\, \d t.
\end{equation}
The second integral on the right-hand side of Eq.~\eqref{eq:final_prop_2} can be evaluated by exchanging the order of integration in $x$ and $t$ and using  the fact that ${\cal H }$ does not depend on time to obtain
\begin{equation}
\int_0^T\int_0^\lambda
\frac{\partial \cal Y}{\partial t}
 \frac{\d \cal H }{\d x } 
\,\d x\, \d t=
\int_0^\lambda  \frac{\d \cal H }{\d x } 
\left(\int_0^T
\frac{\partial \cal Y}{\partial t}
\, \d t\right)\d x =0,
\end{equation}
the last result being a consequence of the periodicity of the function $\cal Y$. We therefore see that the average position of the flagellum, described by the function $\cal H$, does not affect the value of the mean propulsive force generated by the periodic beating.

To complete the calculation for $F_x$, we exploit the fact that $\cal Y$ is a travelling wave, Eq.~\eqref{eq:wavefinal}, so that
\begin{equation}\label{eq:misceqs2}
 \frac{\partial \cal Y}{\partial x} =\frac{\d \cal Y}{\d \eta} ,
\end{equation}
and, together  with Eq.~\eqref{eq:misceqs1},    the propulsive force in Eq.~\eqref{eq:final_prop_2} is given by
 \begin{equation}\label{eq:Fx_final}
F_x 
=-\Gamma c_\perp \frac{\lambda^2}{ T^2}
\int_0^\lambda
[{\cal Y}'(\eta)]^2
\,\d \eta,
\end{equation}
where we have used Eq.~\eqref{eq:Gamma}  to obtain the value of the  Lagrange multiplier. Note that, given Eq.~\eqref{eq:W0}, we   obtain
a   relationship between propulsion and dissipation as
\begin{equation}
F_x =-\Gamma \dot W_0.
\end{equation}

Given that  $\dot W_0 > 0$, we therefore see that there is a direct link between the direction of propagation of the wave and the direction of the propulsive force. Specifically, we see that $\Gamma F_x < 0$. The case $\Gamma > 0$ in Eq.~\eqref{eq:wavefinal} corresponds to a wave travelling in the $+x$ direction, and it results in  minimum negative propulsion, $F_x<0$. By symmetry, the reverse is also true, and a wave travelling in the $-x$ direction ($\Gamma < 0$) leads to maximum propulsion along the $+x$ direction, $F_x>0$. This situation where wave propagation and  swimming occur in a direction  opposite to  one another is observed for all microorganisms using smooth flagella  for locomotion, the direction being reversed for hairy flagella with  
$c_\parallel>c_\perp$~\cite{lighthill75,lighthill76,lp09}.

 \section{Discussion}

The calculus of variations is a mathematical method that has long been  invoked to solve optimisation problems in physical sciences, starting with  classical mechanics.  It also has been useful for many problems  in fluid mechanics, from shape optimisation~\cite{mohammadi2004shape} to recent work using  adjoint methods for flow control and hydrodynamic stability~\cite{kim2007linear,schmid2007nonmodal,luchini2014adjoint}. 
In this paper, we have used the calculus of variations on a long-wavelength model of  flagellar swimming. We have shown that, assuming shapes periodic in space and time,  maximising  the average propulsive force for a fixed amount of energy dissipated in the surrounding fluid breaks the 
right-left symmetry of the problem and leads exactly to travelling-wave solutions. These solutions might feature a non-zero  mean centreline shape, on top of  which a wave propagates at constant speed, and always correspond to locomotion induced in the direction opposite to that of the travelling wave. 
 Importantly, beyond an energetic integral constraint (Eq.~\ref{eq:W0}), the long-wavelength travelling wave shapes are  arbitrary and they are all optimal solutions to the problem.

{The fact that   the optimisation in our paper could be carried out analytically    results from some simplifications in our modelling approach. First, the fluid dynamics forces were included at the level of resistive-force theory, which is known to be valid only in the limit of very slender filaments. Including hydrodynamic interactions beyond the slender limit~\cite{johnson79,johnson80} might modify the details of the optimal kinematics. Recent work on optimal peristaltic motion~\cite{agostinelli2018peristaltic} suggests however that  waves remain optimal  despite a different hydrodynamic setting, suggesting that the predominance  of wave-like motion reflect intrinsic symmetries in the  governing equations.}

{A second modelling choice made in our paper is the  fixed periodicity of the waves. Relaxing this assumption would prevent the cancellation of the boundary terms in the two 
integration by parts in 
Eqs.~\eqref{eq:deltaFx} and \eqref{eq:17}, and the one transforming Eq.~\eqref{eq:Wbefore} into Eq.~\eqref{eq:Wfinal}. However, it is important to note that the transport equation resulting from the  zero first variation of the Lagrangian,   Eq.~\eqref{eq:optPDE},  would remain the same. A more complex setup would consist in not fixing the wavelength (or time period) for the problem, but letting it be part of the optimisation procedure, and it would be instructive to see what optimal periodicity would result from it.}

 {The final modelling assumption in our paper is that of long-wavelength waving.} In their original paper, Pironneau and  Katz~\cite{pironneau74} did not make this assumption. Instead they considered the instantaneous propulsion problem and asked what flagellar kinematics minimised the energy dissipated in the fluid when subject to a mathematically-appropriate (but not intuitive) constraint of   flagellum velocity  projected along the  flagellum tangent and spatially averaged. {They obtained  instantaneously-optimal flagellar kinematics consisting of an apparent sliding of the flagellum, in the body frame, along its instantaneous centreline, suggesting therefore travelling-wave motion.} In contrast with their instantaneous result, we have obtained here exactly travelling waves as the optimal solution to the  time-averaged propulsive problem, and the  waves, which could potentially propagate  on  a curved flagellum,  have a constant wave speed.  
   {Recent numerical work beyond the  long-wavelength approximation  reported convoluted optimal strokes for $N$-link swimmers, suggesting that  high-amplitude motion  can lead to more complex optimal  waving~\cite{alouges2019energy}.}

Finally, like the calculation in Ref.~\cite{pironneau74}, our work was restricted to planar flagella motion, and it would be {enlightening} to also solve the optimisation problem in  the case of a  flagellum   allowed to deform in three dimensions, in particular given the difference between the passive helical rotation of bacterial flagella~\cite{berg00_PT} and the active helical waves of eukaryotes~\cite{chwang71}. 
We have also assumed that the flagellum is infinite and, combined with  its periodicity, this means that hydrodynamic  forces only needed to be resolved along $x$.  Allowing the flagellum to  be  finite 
would require to relax spatial periodicity and lead in general to motion perpendicular to the mean centreline and to   pitching of the cell~\cite{friedrich2010high,neal2020doing}, the magnitudes of which could  also be added as constraints in an extension of our approach to finite flagella.

\section*{Acknowledgements}
I thank Alex Chamolly, Debasish Das, Masha Dvoriashyna, Christian Esparza-L\'opez,  Ray Goldstein, Lyndon Koens and Maria T$\breve{\rm a}$tulea-Codrean 
 for useful discussions and feedback. 
This project has received funding from the European Research Council (ERC) under the European Union's Horizon 2020 research and innovation programme  (grant agreement 682754).

\bibliographystyle{unsrt}

\bibliography{variation}

\end{document}